\documentclass{article}

\title{Registers}
\author{
Paul Vit\'{a}nyi
 \\
CWI and University of Amsterdam
}

\newcommand{\ra}{\longrightarrow}

\date{}
\begin{document}

\maketitle
\section{Synonyms}
Wait-free registers, wait-free shared variables, asynchronous communication
hardware.
\section{Problem Definition} \label{sec1}
Consider a system of asynchronous processes that communicate among themselves
by only executing read and write operations on a set of 
shared variables (also known
as shared {\em registers}).
The system has no global clock or other synchronization primitives.
Every shared variable is associated with a process (called {\em owner\/}) which writes it and
the other processes may read it.
An execution of a write (read) operation on a shared variable will
be referred to as a {\em Write\/}  ({\em  Read\/}) on that variable.
A Write on a shared variable puts a value from a pre-determined finite domain into the
variable, and a Read   reports a value from the domain.
A process that writes (reads) a variable is called a {\em writer\/} ({\em reader\/}) of the
variable.

We want to construct shared variables in which 
the following two properties hold.
(1)~Operation executions are not necessarily atomic, 
that is, they are not indivisible
but rather consist of atomic sub-operations,
and (2)~every operation finishes its execution 
within a bounded number of its
own steps, irrespective of the presence of other operation executions and
their relative speeds. That is, operation executions are {\em wait-free}.
These   two properties give rise to a classification of shared variables,
depending on their output characteristics.
Lamport \cite{lamp86} distinguishes three 
categories for 1-writer shared variables,
using a precedence relation on operation executions defined as follows:
for operation executions $A$ and $B$, $A$ {\em precedes\/} $B$, denoted $A\ra
B$, if $A$ finishes before $B$ starts; $A$ and $B$ {\em overlap\/} if neither
$A$ precedes $B$ nor $B$ precedes $A$.
In 1-writer variables, all the Writes are totally ordered by ``$\ra$''.
The three categories of 1-writer shared variables defined by Lamport are the following.
\begin{enumerate}
\item A {\em safe\/} variable is one in which a Read not overlapping any Write
returns the most recently written value.
A Read that overlaps a Write may return any value from the domain of the variable.
 
\item A {\em regular\/} variable is a safe variable in which
a Read that overlaps one or more Writes returns either the value
of the most recent Write preceding the Read or of one of the overlapping
Writes.
 
\item An {\em atomic\/} variable is a regular variable in which
the Reads and Writes behave as if they occur in some total order
which is an extension of the precedence relation.
\end{enumerate}
 
A shared variable is {\em boolean\/}\footnote{Boolean variables are referred to as {\em bits}.} 
or {\em multivalued\/} depending upon whether it can hold only two or more than two values.
A {\em multiwriter} shared variable is one that can be written and read
(concurrently) by many processes. If there is only one writer and more
than one reader it
is called a {\em multireader} variable.

\section{Key Results}
In a series of papers 
starting
in 1974, for details see \cite{HaVi02}, Lamport
explored various notions of concurrent reading and writing of shared
variables culminating in the seminal 1986 paper \cite{lamp86}. It formulates
the notion of wait-free implementation of an atomic multivalued
 shared variable---written
by a single writer
and read by (another) single reader---from safe 1-writer 1-reader
2-valued shared variables, being mathematical versions of physical 
{\em flip-flops}, later optimized in \cite{tromp89}.
Lamport 
did not consider constructions
of shared variables with more than one writer or reader.

 Predating the Lamport 
paper, in 1983 Peterson
\cite{pete83} published an ingenious wait-free construction of an atomic
1-writer, $n$-reader $m$-valued atomic shared variable from
$n+2$ safe 1-writer $n$-reader $m$-valued registers, $2n$ 
1-writer 1-reader 2-valued
atomic shared variables, and 2 1-writer $n$-reader 
2-valued atomic shared
variables. He presented also a proper notion of wait-freedom property.
In his paper, Peterson didn't tell how to construct the $n$-reader
boolean atomic variables from flip-flops, while Lamport mentioned the open
problem of doing so, and, incidentally, uses a version
of Peterson's construction to bridge the algorithmically demanding
step from atomic shared bits to atomic shared multivalues.
Based on this work, N. Lynch, motivated by concurrency control of
multi-user data-bases, posed around 1985 the question of how to
construct wait-free multiwriter atomic variables from
1-writer multireader atomic variables. Her student Bloom
\cite{bloom87} found in 1985 an elegant 2-writer construction, which,
however, has resisted generalization to multiwriter. Vit\'anyi and Awerbuch
\cite{vit86} were the first to define and explore the complicated notion of 
wait-free constructions of general
multiwriter atomic variables, 1986. They presented
a proof method, an unbounded solution 
from 1-writer 1-reader atomic variables, and a bounded solution
from 1-writer $n$-reader atomic variables.
The bounded solution 
turned out
not to be atomic, 
but only achieved regularity
(``Errata'' in \cite{vit86}).
The paper introduced important notions and techniques in the
area, like (bounded) vector clocks, and identified open problems
like the construction of atomic wait-free  bounded multireader shared variables
from flip-flops, and atomic wait-free bounded multiwriter shared variables
from the mulireader ones. 
Peterson who had been working on the multiwriter problem
for a decade,
together with Burns, tried in 1987 to eliminate the error in the
unbounded construction of \cite{vit86} 
retaining the idea of vector clocks, but replacing the
obsolete-information tracking technique
by repeated scanning as in \cite{pete83}. The result 
\cite{pete87} was found to be erroneous in the technical
report 
(R. Schaffer, On the correctness of atomic multiwriter registers,
	Report~MIT/LCS/TM-364, 1988).
Neither the 
re-correction in Schaffer's Tech Report, nor the claimed re-correction by
the authors of \cite{pete87} has appeared in print.
Also in 1987 there appeared at least five purported solutions for the
implementation of 1-writer $n$-reader atomic shared variable from
1-writer 1-reader ones: \cite{kiro87,burn87,sing87} (for the others
see \cite{HaVi02})
of which \cite{burn87} was shown to be incorrect
(S. Haldar, K. Vidyasankar, 
{\em ACM Oper. Syst. Rev}, 26:1(1992), 87--88)
and only \cite{sing87} appeared in journal version.
The paper \cite{li89},
intially a 1987
Harvard Tech Report, 
 resolved all multiuser constructions
in one stroke:
it constructs a bounded $n$-writer $n$-reader
(multiwriter) atomic variable from $O(n^2)$
1-writer 1-reader safe bits, which is optimal, and $O(n^2)$ bit-accesses
per Read/Write operation which is optimal as well.
It works by making the unbounded solution of \cite{vit86}
bounded, using a new technique, achieving a robust proof of correctness. 
``Projections'' of the construction give specialized constructions
for  the implementation of $1$-writer $n$-reader (multireader) atomic 
variables from $O(n^2)$  $1$-writer $1$-reader ones using $O(n)$ bit
accesses per Read/Write operation, and for the implementation
of $n$-writer $n$-reader (multiwriter) 
atomic variables from $n$  1-writer $n$-reader (multireader) ones. 
The first ``projection'' is optimal, while the
last ``projection'' may not be optimal since it uses $O(n)$
control bits per writer while only a lower bound of $\Omega (\log n)$
was established. Taking up this
challenge, the construction in \cite{isra92a} claims to achieve this
lower bound. 

{\bf Timestamp system:}
In a multiwriter shared variable it is only required that every process keeps track
of which process wrote last. There arises the general question whether every process
can keep track of the order of the last Writes by all processes.  
A. Israeli and M. Li were attracted to the area by the work in \cite{vit86},
and, in an important paper \cite{isra93}, they raised and solved the question
of the more general and universally useful
notion of bounded timestamp system to track the order
of events in a concurrent system. 
In a timestamp system
every process owns an {\em object\/}, 
an abstraction of a set of shared variables.
One of the requirements of the system is to determine the temporal order in
which the objects are written.
For this purpose, each object is given a {\em label\/} 
(also referred to as {\em
timestamp}) which indicates the latest (relative) time when it has been 
written by its owner process.
The processes assign labels to their respective objects in such a way that the 
labels reflect the real-time order in which they are written to. 
These systems must support two operations, namely {\em labeling\/} and {\em
scan}.
A labeling operation execution (Labeling, in short) assigns a new label to an
object, and a scan operation execution (Scan, in short) enables a process to
determine the ordering in which all the objects are written,
that is, it returns a set of labeled-objects ordered temporally.
We are concerned with those systems where operations can be
executed {\em concurrently}, in an overlapped fashion.
Moreover, operation executions must be {\em wait-free}, that is, each
operation execution will take a bounded
number of its own steps (the number of
accesses to the shared space), irrespective of the presence of other operation
executions and their relative speeds.
Israeli and Li \cite{isra93} 
constructed a bit-optimal bounded timestamp system 
for {\em sequential} operation executions.
Their sequential timestamp
system was published in the above journal reference, 
but the preliminary concurrent
timestamp system in the conference proceedings, of which a more
detailed version has been circulated in manuscript form,
has not been published in final form. 
The first generally
accepted solution of the 
 {\em concurrent} case of bounded 
timestamp system is 
due to Dolev and Shavit
\cite{dole89}.
Their construction is of the type as in \cite{isra93}
and uses shared variables of size $O(n)$, where $n$ is the
number of processes in the system.
Each Labeling requires $O(n)$ steps, and each Scan $O(n^2\log n)$ steps.
(A `step' accesses an $O(n)$ bit variable.)
In \cite{HaVi02} 
the unbounded construction of \cite{vit86} is corrected and extended 
to obtain an efficient version of the more general
notion of bounded concurrent timestamp system.

\section{Applications}

Wait-free registers are, together with message-passing systems,
the primary interprocess communication method in distributed computing
theory. They form the basis of all constructions and protocols, as can
be seen in the textbooks.
Wait-free constructions of concurrent timestamp systems (CTSs, in short)
have been 
shown to be a powerful tool for solving concurrency control problems such as
various types of mutual exclusion,
multiwriter multireader shared variables
\cite{vit86}, and probabilistic consensus, 
by synthesizing a ``wait-free
clock'' to sequence the actions in a concurrent system.
For more details see \cite{HaVi02}.

\section{Open Problems}
There is a great deal of work in the direction of register constructions
that use less constituent parts, or simpler such parts, or parts that
can tolerate more complex failures, than previous
constructions referred to above. Only, of course, if the latter constructions
were not yet optimal in the parameter concerned. Further directions are
work on wait-free higher-typed objects, as mentioned above, hierarchies
of such objects, and probabilistic constructions.
This literature is too vast and diverse than can be surveyed here.

\end{document}